\documentclass{mem}
\usepackage{natbib}\usepackage{txfonts}\usepackage{balance}
\usepackage{graphicx}
\usepackage{txfonts}
\usepackage[a4paper]{hyperref}
\idline{79}{1}
\begin{document}

\def\dexkpc{$\textrm {dex kpc}^{-1}$ }
\def\teff{$T\rm_{eff}$ }
\def\kms{$\mathrm {km s}^{-1}$ }
\def\logg{{\bf$log~g$} }
\def\vt{v$_{t} $}

\title{Cepheids as tracers of the metallicity gradient across the Galactic disk$^{*}$
   \thanks{Based on observations obtained with ESPADONS at the Canada-France-Hawaii Telescope (CFHT) 
and on observations collected with FEROS at the ESO/MPI 2.2
m. telescope at ESO, La Silla. }
}

\author{
B. \,Lemasle\inst{1,2} 
\and A. \,Piersimoni\inst{3}
\and S. \,Pedicelli\inst{4}
\and G. \,Bono\inst{4,5}
\and P. \,Fran\c cois\inst{2}
\and F. \,Primas\inst{4}
\and M. \,Romaniello\inst{4}
}

  \offprints{B. Lemasle}

\institute{
Universit\'e de Picardie Jules Verne, Facult\'e des Sciences, 33 Rue Saint-Leu, 80039 Amiens Cedex 1, France
\and
Observatoire de Paris-Meudon, GEPI, 92195 Meudon Cedex, France
\and
INAF - Osservatorio Astronomico di Collurania, via M. Maggini, 64100 Teramo, Italy
\and
European Southern Observatory (ESO), Karl Schwarzschild-Strasse 2, 85748 Garching bei Muenchen, Germany
\and
INAF - Osservatorio Astronomico di Roma, via Frascati 33, 00040 Monte Porzio Catone, Italy\\
\email{bertrand.lemasle@etud.u-picardie.fr}
}

\authorrunning{Lemasle et al.}

\titlerunning{Cepheids as tracers of the Galactic disk metallicity gradient.}

\abstract{

We present iron abundance measurements, based on high resolution 
spectroscopy, and accurate distance determinations, based on near infrared 
photometry, for 34 Galactic Cepheids. The new data are used to constrain 
the Galactic iron abundance gradient in the outer disk, namely from 10 to 
14 kpc. We confirm the flattening of the gradient toward the outer disk.
In this region we also found an increase in the metallicity dispersion.
Current data do not support the occurrence of a jump in the metallicity 
gradient for Galactocentric distances of the order of 10-12 kpc.

\keywords{Stars: abundances -- Stars: supergiants -- Galaxy: abundances -- Galaxy: evolution}
}
\maketitle{}

\section{Introduction}

Galactic abundance gradients are a fundamental input for chemodynamical evolutionary 
models, since they are the observables typically adopted to validate predictions. 
Among the different tracers used to determine the Galactic gradients: HII regions, 
Open Clusters, O/B-type stars, Planetary Nebulae, we chose the Cepheids.  The 
Cepheids present several advantages when compared with other tracers:
{\em i)} they are excellent distance indicators, which is a strong positive feature 
to provide accurate Galactocentric distances; 
{\em ii)} they are also bright enough to allow the study of the gradient over a large 
range of Galactocentric distances;
{\em iii)} they present a large set of well defined absorption lines, therefore, accurate 
and precise abundances of many heavy elements can be provided.    

Abundance gradients were discovered first in six external galaxies from HII regions by 
\citet{Sea71} but their occurrence in the Galaxy was a controversial issue. Indeed, 
some investigations were in favor of Galactic gradients \citep{D'odo76,Jan79}, 
while others found no correlation between metallicity and distance \citep{Cle73,Jen75}.

Even though the presence of Galactic abundance gradients seems widely
accepted, the empirical estimates of the slopes are still lively
debated.  By using HII regions, \citet{Vil96} found a slope of $-0.02$ \dexkpc,
but different authors using the same tracers suggest slopes ranging from $-0.039$ \dexkpc 
\citep{Deh00} to $-0.065$ \dexkpc \citep{Aff97}. On the other hand, 
the use of B-type stars gives slopes ranging from $-0.042$ \dexkpc \citep{Daf04} 
to $-0.07$ \dexkpc \citep{Gum98}. The slopes based on Planetary Nebulae range
from $-0.05$ \dexkpc \citep{Cos04} to $-0.06$ \dexkpc \citep{Mac99} to the lack 
of a Galactic metallicity gradient \citep{Sta06} . The slopes based on old Open 
Clusters still show a large spread. By adopting a sample of 40 clusters distributed 
between the solar circle and $R_G\simeq14$ kpc, \citet{Frie02} found a slope of 
$-0.06$ \dexkpc. More recently, \citet{Car07} using new accurate metal abundances 
for five old open clusters located in the outer disk together with the sample 
adopted by Friel et al. (2002) found a much shallower global iron gradient, 
namely $-0.018\pm0.021$ \dexkpc.  
The global iron slopes based on Cepheids are very homogeneous, and indeed 
dating back to the first estimates by \citet{Har81,Har84}, who found a slope 
of $-0.07$ \dexkpc, the more recent estimates provide slopes ranging from 
$-0.06$ \dexkpc \citep{And02a,And02b,And02c,Luck03,And04,Luck06} to  
$-0.07$ \dexkpc \citep{Lem07}. 

Concerning the shape of the gradient, the situation is even more complicated. 
The hypothesis of a linear gradient is very disputed. Several authors favor 
a flattening of the gradient beyond 10-12 kpc: \citet{Vil96} (HII regions), 
\citet{Twa97} (Open Clusters), \citet{And04} (Cepheids), \citet{Cos04} 
(Planetary Nebulae). This flattening is also well reproduced by models 
\citep{Ces07}.  In a recent paper, \citet{Yong06} brought another feature: 
the flattening may occur with two basement values, the first one at $-0.5$ dex, 
a lower value than previous studies, and the second one at $-0.8$ dex, for which 
they suspected the possibility of a merger event. On the other hand, independent 
investigations do not show a clear flattening toward the outer disk, like 
\citet{Roll00} (O,B stars) and \citet{Deh00} (HII regions). Morever, a change in 
the slope in the direction of the inner disk was suggested by \citet{And02b} 
in a study based on a limited sample of Cepheids.
In this context it is worth mentioning that the different tracers adopted to 
estimate the global iron gradient present a handful of objects in the outer 
disc, i.e. at $R_G\ge12$ kpc. Finally, we mention that it has also been 
suggested by \citet{Twa97}, using Open Clusters, and by \citet{And04}, using 
Cepheids, a jump in the metallicity gradient at $R_G\sim10-12$ kpc, with 
the iron abundance sharply decreasing by $\approx-0.2$ dex. 
Here we investigate the shape of the gradient in the outer disk and focus 
our attention on the possible jump in metallicity.

\section{Observations and Methods}

Our data sample includes high resolution spectra of 34 Galactic Cepheids. 
A large fraction of them (28) were collected with ESPADONS at CFHT, whereas 
six were collected with FEROS at 2.2m ESO/MPG telescope \citep{Kau99}. 
Spectra were reduced using either the FEROS package within MIDAS or 
the Libre-ESpRIT software \citep{Don97,Don06}. The data analysis was 
already described in \citet{Lem07}. Suffice it here to briefly mention 
the main steps of the reduction strategy. The accurate determination 
of effective temperatures is a critical point in the abundance determination. 
They are only spectroscopically determined, by using the method of line depth 
ratios (LDR), described in \citet{KovGor00}. As it relies only on spectroscopy, 
it is independent of interstellar reddening.  These authors proposed 32 analytical 
relations for determining \teff  from LDR of weak, neutral metallic lines.

The other atmospheric parameters (surface gravity \logg, microturbulent velocity 
\vt) are determined by imposing the ionization balance between FeI and FeII with 
the help of curves of growth: iterations on \logg  and \vt~ are repeated until 
the best match with the same curve of growth is reached. 
Abundances are then calculated with an atmospheric model from \citet{Edv93} based 
on the following assumptions: parallel plane stratification, hydrostatic equilibrium 
and LTE. We adopted the scaled-solar chemical abundances suggested by \citet{Gre96}. 
The final internal accuracy in the abundance determination is of the order of 0.1 dex.

Absolute distances were estimated using near infrared photometry ($J,H,K-$band) 
from the 2MASS catalog. The mean magnitude of these objects was estimated using 
the template light curves provided by \citet{Sos05}, together with the $V-$band 
amplitude and the epoch of maximum available in the literature.
For ST Tau and AD Gem the data from \citet{Bar97} were also used, while for 
HW Pup the data from \citet{Sche92} have been joined to the 2MASS ones.
As templates are only for fundamental modes pulsators, the mean near infrared
magnitudes for first overtone pulsators are only the single 2MASS measurement.
The Cepheid absolute magnitudes were computed using the near infrared 
period-luminosity relations recently provided by \citet{Per04}, 
with an LMC distance modulus of 18.50 mag. Pulsation periods and reddenings are from the 
Fernie's database \citep{Fer95} and period of first overtone pulsators 
were fundamentalized. We assumed a Galactocentric distance of 8.5 kpc 
for the sun \citep{Fea97}.

\section{Results and final remarks}

Current iron abundances when compared with previous measurements, 
mainly from the systematic investigation by Andrievsky et al., 
show a very good agreement. For 17 stars the abundance difference 
is very small ($<$0.1 dex) and for 6 stars it is $\sim0.1$ dex. 
We cannot confirm previous determinations only for 4 stars, for 
which discrepancies can reach $0.3$ dex.  To improve the sampling 
along the Galactic disk, we added to our new data 30 Cepheids 
from \citet{Lem07}, 52 Cepheids from Andrievsky's sample and 
11 Cepheids from \citet{Mot07}, for which both metallicities 
and infrared photometry were available. The complete sample includes 
127 Cepheids with distances based on homogeneous infrared photometry.
Among them 27 have Galactocentric distances in the 10-12 kpc range 
and 13 are located beyond 12 kpc.

\begin{figure}[!ht]
\begin{center}
\includegraphics[angle=90,width=7cm]{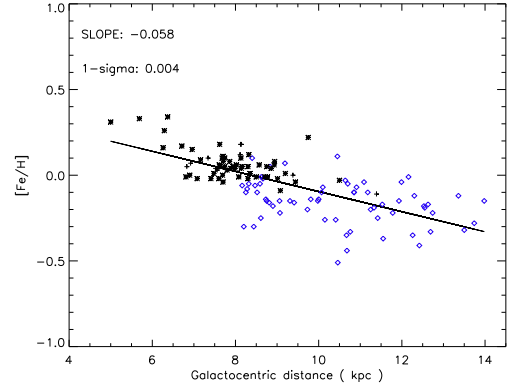}
\caption{Galactic radial abundance gradient. The solid line shows the linear regression whose 
slope and $1-\sigma$ error are labeled.
Diamonds mark our data, while asterisks show data from Andrievsky et al. and plus signs display 
data from Mottini et al.}
\label{fig:full_data}
\end{center}
\end{figure}

We first estimated the Galactic gradient for the whole Cepheid sample. 
Data plotted in Fig. \ref{fig:full_data} bring forward the following 
findings: 

\noindent 
$\bullet$ The iron gradient shows no evidence of a gap over the 
Galactocentric distances covered by the current sample. \\
$\bullet$ The iron gradient is flattening in the outer disk.\\
$\bullet$ The spread in metallicity is larger in the outer disk, 
possibly due to local inhomogeneities.\\

\begin{figure}[!h]
\begin{center}
\includegraphics[angle=90,width=7cm]{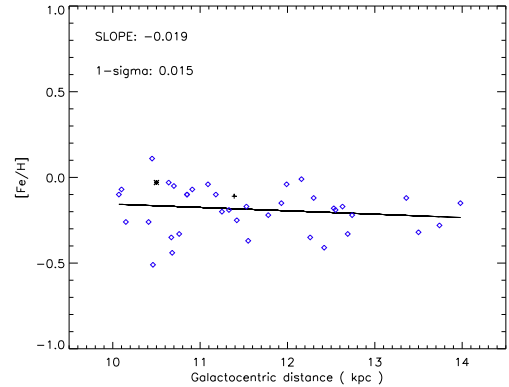}
\caption{Same as Fig.1, but for Cepheids with Galactocentric distances 
ranging from 10 to 15 kpc range.}
\label{fig:outer_data}
\end{center}
\end{figure}

The slope of the linear iron gradient over the entire sample is shallower,  
$-0.058$ \dexkpc, than previously estimated by \citet{Lem07}, but the new 
sample extends toward the outer disk, where the flattening occurs. This 
slope is in very good agreement with previous determinations from 
\citet{Frie02} (Open Clusters) and \citet{Luck06} (Cepheids). 
The flattening of the gradient in the outer disk becomes even more 
evident in Fig. \ref{fig:outer_data}. We estimated the slope by using 
Cepheids located between 10 and 15 kpc and we found $-0.019$ \dexkpc. 
This estimate is in excellent agreement with the slope found by 
\citet{Car07} using old Open Clusters, i.e. $-0.018\pm0.02$ 
\dexkpc. Moreover and even more importantly, the two independent 
estimates suggest, within the errors, a very similar metallicity 
value in the outer disc, namely $[Fe/H]=-0.3,-0.4$.  

Several reasons can be invoked to explain the lack of the metallicity gap 
in our sample. The first one was suggested by \citet{Luck06}, who pointed 
out that their sample, as well as the one by \citet{Twa97} was gathering 
stars lying approximately at the same Galactic longitude, respectively 
190-250 and 130-260 degrees. Moreover, the uncertainties on the distance 
determination increase with the star distances (especially when the distance 
is estimated using optical photometry) and could introduce artifact in the 
distribution. The use of IR photometry and robust primary distance indicators 
presents several undisputed advantages.
Note that the outer disk in strongly undersampled, therefore, the selection 
of the targets among the few available Cepheids can introduce a bias that 
can only be removed by increasing the sample size in this region.
Finally, we note that the lack of a gap in our iron gradient estimate might 
be due to the restricted range of longitudes covered by our sample. 
This relevant point deserves further investigations.

\section{Summary}

High resolution spectra and accurate distance determinations for 34 Cepheids 
provided the opportunity to investigate the iron gradient across the Galactic 
disc. Preliminary results indicate that:\\
$\bullet$ The Galactic iron gradient presents a linear trend and the slope is 
$\approx-0.06$ \dexkpc.\\
$\bullet$ The shape of the gradient is more accurately described by a bimodal 
distribution, with a higher slope in the central region and a flattening of 
the gradient in the outer disk. In this region, the spread in metallicity is 
higher, possibly due to local inhomogeneities.  A third zone can be considered 
in the inner disk, with a break in the slope near 7 kpc, but this region was 
not explored in our study.\\
$\bullet$ Current data show no evidence of a jump in the iron gradient for 
$R_G\sim 10-12$ kpc.


\bibliographystyle{aa}

\end{document}